\documentclass[prb,twocolumn,showpacs,preprintnumbers,amsmath,amssymb,superscriptaddress]{revtex4}
\usepackage{graphicx}
\usepackage{dcolumn}
\usepackage{bm}

\begin{document}

\title{Fine structure of ``zero-mode'' Landau levels in HgTe/HgCdTe quantum wells}
\author{M.~Orlita}\email{milan.orlita@lncmi.cnrs.fr}
\altaffiliation{also at Institute of Physics, Charles University, Ke Karlovu 5,
CZ-121~16 Praha 2, Czech Republic \& Institute of Physics, v.v.i., ASCR,
Cukrovarnick\'{a} 10, CZ-162 53 Praha 6, Czech Republic} \affiliation{Laboratoire National des Champs Magn\'etiques Intenses, CNRS-UJF-UPS-INSA, 25, avenue des Martyrs, 38042 Grenoble, France}

\author{K.~Masztalerz}
\altaffiliation{present address: Institute of Experimental
Physics, University of Warsaw, Ho\.{z}a 69, PL 00-681 Warsaw,
Poland} \affiliation{Laboratoire National des Champs Magn\'etiques Intenses, CNRS-UJF-UPS-INSA, 25, avenue des Martyrs, 38042 Grenoble, France}
\author{C.~Faugeras}
\affiliation{Laboratoire National des Champs Magn\'etiques Intenses, CNRS-UJF-UPS-INSA, 25, avenue des Martyrs, 38042 Grenoble, France}
\author{M.~Potemski}
\affiliation{Laboratoire National des Champs Magn\'etiques Intenses, CNRS-UJF-UPS-INSA, 25, avenue des Martyrs, 38042 Grenoble, France}

\author{E.~G.~Novik}
\affiliation{Physikalisches Institut (Lehrstuhl f\"{u}r Experimentelle Physik
III), Universit\"{a}t W\"{u}rzburg, D-97074 W\"{u}rzburg, Germany}
\author{C.~Br\"{u}ne}
\affiliation{Physikalisches Institut (Lehrstuhl f\"{u}r Experimentelle Physik
III), Universit\"{a}t W\"{u}rzburg, D-97074 W\"{u}rzburg, Germany}
\author{H.~Buhmann}
\affiliation{Physikalisches Institut (Lehrstuhl f\"{u}r Experimentelle Physik
III), Universit\"{a}t W\"{u}rzburg, D-97074 W\"{u}rzburg, Germany}
\author{L.~W.~Molenkamp}
\affiliation{Physikalisches Institut (Lehrstuhl f\"{u}r Experimentelle Physik
III), Universit\"{a}t W\"{u}rzburg, D-97074 W\"{u}rzburg, Germany}
\date{\today}

\begin{abstract}
HgTe/HgCdTe quantum wells with the inverted band structure have
been probed using far infrared magneto-spectroscopy. Realistic
calculations of Landau level diagrams have been performed to
identify the observed transitions. Investigations have been
greatly focused on the magnetic field dependence of the peculiar
pair of ``zero-mode'' Landau levels which characteristically split
from the upper conduction and bottom valence bands, and merge
under the applied magnetic field. The observed avoided crossing of
these levels is tentatively attributed to the bulk inversion
asymmetry of zinc blend compounds.
\end{abstract}

\pacs{71.55.Gs, 71.70.Di, 73.20.At, 85.75.-d}

\maketitle

\section{Introduction}

Experimental finding of the quantum spin Hall effect (QSHE) in
HgTe/CdTe quantum wells\cite{KonigScience07} (QWs) affirmed the
previous theoretical predictions\cite{BernevigScience06} and has
had a relevant impact on further development of the research on
two- and three-dimensional (2D and 3D) topological
insulators. Remarkably, the origin of the
topological insulator phase (bulk gapped insulator with gapless
conducting states at the edges or surfaces) turns out to be just
the peculiar band structure, characteristic of certain class of
semiconductors with an appropriate strength of spin-orbit
interaction.\cite{QiPT10,HasanRMP10}

The 2D archetype of this phase appears in HgTe quantum wells in
the regime of the so-called inverted band structure (semiconductor
with a gap between the upper $p$-type and lower $s$-type energy
band). The inherent property of such bands is their characteristic
behavior under the applied magnetic field, i.e., the appearance of
the particular pair of Landau levels (LLs) which distinctly split
from the upper and lower energy band. Progressively with the
magnetic field, those two Landau levels merge and eventually cross
at certain field $B_c$, above which the zero-field topologically
insulating phase is transformed into the common quantum Hall
insulator.

The origin of this distinct pair of LLs and its possible
interesting physics becomes apparent when using a modified
$4\times4$ Dirac-type Hamiltonian for the approximate description
of the Fermi level-vicinity electronic states. In case of HgTe
QWs, the off-diagonal terms of the massless Dirac Hamiltonian have
to be completed by the diagonal (dispersive) mass terms. Nevertheless, the
above-mention particular pair of Landau levels still appears due to
the off-diagonal, linear in particle momentum terms. It is a
reminiscence of the characteristic, field-independent zero-mode
(zero-index) Landau levels of masless Dirac fermions, such as, for
example, those found in
graphene.\cite{FuchsPRL07,JiangPRL07transport,ArikawaPRB08,JungPRB09,ZhangPRL10}
In HgTe QWs, these levels are split in energy due to the mass term
and their positions change monotonically as a function of the
width of the quantum well. In case of inverted bands, this
splitting becomes negative what intuitively accounts for merging
of zero-mode LLs in a magnetic field. Notably, the zero-mode
Landau levels in HgTe are spin polarized (even if the bare Zeeman
splitting is ignored) similarly like the zero-mode levels are
pseudo-spin polarized in graphene, i.e., with charge localized on either of two
triangular sublattices.

As reported before,\cite{KonigScience07} the magnetic field
evolution of the zero-mode LLs in HgTe QWs with the inverted band
structure is at the origin of the field-driven
insulator-metal-insulator phase transition, characteristic of these
systems. A simple scenario of crossing of those levels at a
critical field $B_c$ has been sufficient to account for the
magneto-transport data, but a possibility of a weak anti-crossing
effect has also been considered.\cite{KonigScience07,KonigJPSJ08}

In this paper, we inspect more closely the field evolution of
these ``zero-mode'' Landau levels using a suitable experimental tool
of Landau level spectroscopy. Essentially, we find that the
zero-mode Landau levels in HgTe quantum wells with the inverted
band structure do not cross, but instead display the effect of the
avoided crossing. This effect might be due to the bulk inversion
asymmetry (BIA)\cite{DresselhausPR55} which is inherently present
in zinc-blend crystals but habitually neglected in band structure
calculations of HgTe quantum wells.\cite{NovikPRB05} A more
speculative interpretation of our experimental finding would
consist of refereeing to the extended investigations of the
zero-mode Landau levels in graphene\cite{JungPRB09} and pointing
out the electron-electron interaction as a possible source of
opening of a gap within the otherwise double degenerate Landau
level at the critical field $B_c$.

\section{Experimental details}

We have studied two different samples, each containing a
8~nm-wide, [001]-oriented HgTe quantum well embedded in-between
Hg$_{0.3}$Cd$_{0.7}$Te barriers. The band structure of QWs in both
samples is inverted. Since we work with relatively narrow QWs, the
first electron-like subband (E$_1$) is above the second heavy-hole
subband (HH$_2$). This is in contrast to the case of structures
used in previous magneto-optical
studies.\cite{SchultzJCG98,SchultzPRB98} The sample denoted here
as $A$ was intentionally undoped but its residual $p$- or
$n$-doping at the level up to $10^{11}$~cm$^{-2}$ is not excluded.
The QW in the sample B was symmetrically doped with iodium donors,
separated from the well by 40~nm-wide barrier spacers. The density
of 2D electron gas (2DEG), $n=4.2\times10^{11}$~cm$^{-2}$, has
been determined from magneto-transport measurements on a parent
sample (not discussed in this paper). This value served us as an
input parameter for calculations of the band structure and the
resulting energy ladder of LLs in this sample.


To measure the infrared transmittance, the sample was exposed to
the radiation of a globar, which was analyzed by a Fourier
transform spectrometer and delivered to the sample via light-pipe
optics. The transmitted light was detected by a composite
bolometer which operated at $T=2$~K and was placed directly below
the sample. The magneto-optical spectra were measured using either
a superconducting coil ($B\leq 13$~T, applied spectral resolution
of 0.5~meV) or a resistive solenoid (magnetic fields up to 30~T, applied
spectral resolution of 1~meV). All the spectra presented here were
normalized by the sample transmission at $B=0$.

\begin{figure}[b]
\scalebox{0.43}{\includegraphics*{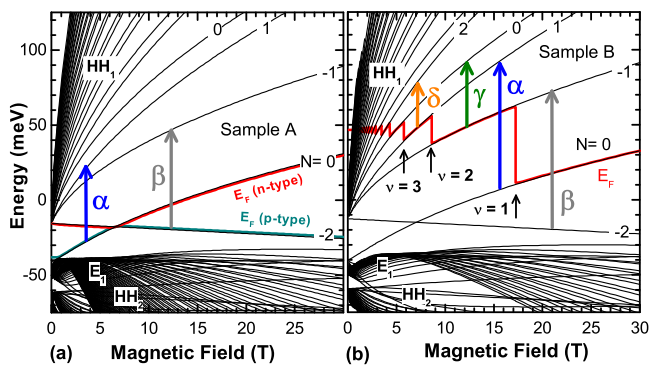}}
\caption{\label{LLoccupation} Calculated dispersion of Landau
levels for both studied samples. The expected dominant absorption
transitions are denoted by arrows and Greek letters. Whereas the
$\alpha$ and $\beta$ lines should be observable in both samples,
the $\gamma$ and $\delta$ transitions are directly related to the
presence of 2DEG in the sample B. The Fermi level is shown in both
cases by a solid line. Two lines for $E_F$ are plotted for the
sample A, as the type of the residual doping is not known.}
\end{figure}

\section{Landau level spectrum in HgTe/HgCdTe quantum wells}

\begin{figure}
\scalebox{1.2}{\includegraphics*{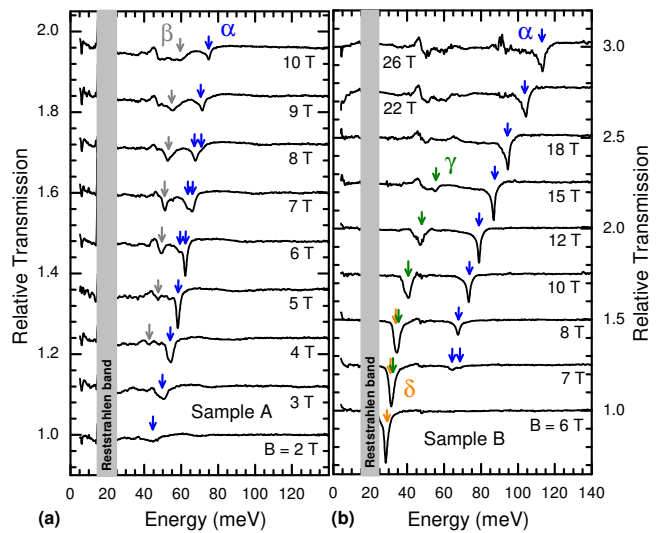}} \caption{\label{SPKT}
Transmission spectra of both investigated samples in a relevant
range of magnetic fields. The notation of the observed lines
follows Ref.~\onlinecite{SchultzPRB98} and is in agreement with
the transitions' assignment used in Fig.~\ref{LLoccupation}.
Samples are completely opaque in the range of the HgTe/HgCdTe
reststrahlen band (14-25~meV).}
\end{figure}

\begin{figure*}
    \begin{minipage}{0.62\linewidth}
      \scalebox{0.37}{\includegraphics{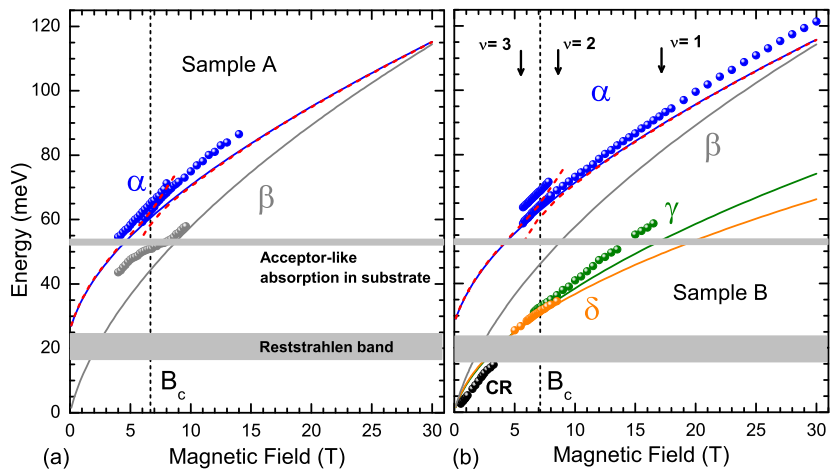}}
    \end{minipage}\hfill
    \begin{minipage}{0.32\linewidth}
      \caption{\label{FanChart} (color online) Fanchart of inter-LL transitions in both
investigated samples. Solid lines show the expected
single-particle excitation energies following the calculation
presented in Fig.~\ref{LLoccupation}. In both samples, the
experimentally traced $\alpha$ transition shows a double component
character in the vicinity of the crossing fields $B_c$. The dashed
lines represent the expected position of the $\alpha$ transition
taking account the mixing of $N=0$ and $-2$ Landau levels,
calculated according to K\"{o}nig \emph{et
al.}\cite{KonigJPSJ08} for the phenomenological coupling
parameter $\Delta=2$~meV.}
    \end{minipage}
\end{figure*}

The detailed LL spectrum of HgTe/HgCdTe QWs is in general derived
from the Kane's band structure model and the resulting $8\times8$
Hamiltonian.~\cite{Pfeuffer-JeschkePhD00,NovikPRB05} This, at
$B>0$, implies solving a set of eight coupled differential
equations for a given LL index $N$, which results in eight
independent solutions for $N>0$ (four pairs of spin-split LLs).
Importantly, only 7, 4 and 1 non-trivial solutions are obtained
for low indices $N = 0,-1$ and -2, respectively. A single, $N=-2$
LL has a purely heavy-hole character ($S=-3/2$) and its energy
decreases nearly linearly with $B$. This level together with one
of the characteristic solutions for $N=0$ represent our zero-mode
LLs which we have identified within a simplified approach of the
Dirac-type Hamiltonian in the introductory part of this paper.

In calculations of LL diagrams of our two QW structures, we have
applied a general scheme of the Kane model but neglected the BIA
effect. Such an approximation implies that, for any QW structure
in the inverted regime, the two zero-mode LLs (or the
characteristic $N=-2$ and $N=0$ LLs) simply cross each other at a
given (well width dependent) magnetic field $B_c$. These
characteristic levels and their crossing can be easily recognized
in Fig.~\ref{LLoccupation} for the case of our samples.

Figure~\ref{LLoccupation} illustrates the details of our LL
calculations in the relevant energy range of the proximity of the
E$_1$, HH$_1$ and HH$_2$ subbands and depicts the transitions
which are expected to be observed in the experiment. Optically
active inter-LL transitions follow the usual $\Delta N=\pm 1$ (for
$\sigma^{\pm}$ circularly polarized light) selection rules imposed
by the electric dipole approximation. The oscillator strength of
the transitions depends also on the overlap of LL wavefunction in
the $z$-direction, what is, however, not essential for the present
considerations. Concluding the previous studies of
HgTe QWs,\cite{SchultzJCG98,SchultzPRB98} we expect the
magneto-optical response of our structures to be dominated by
transitions between LLs with low indices. Those transitions are
marked in Fig.~\ref{LLoccupation} with small Greek letters, in
accordance to the notation of Schultz~\emph{et
al.}\cite{SchultzPRB98}  Transition $\alpha$ accounts for the
expected dominant interband LL transition whereas $\beta$,
$\gamma$ and $\delta$ denote the cyclotron resonance (CR) like
transitions between the adjacent LLs of the same (HH$_1$) subband.

\section{Results and Discussion}

Typical transmission spectra measured at a few selected values of
the magnetic field are plotted in Fig.~\ref{SPKT}. The
corresponding, energy versus magnetic field LL diagrams of the
observed transitions are shown in Fig.~\ref{FanChart}. Inspection
of the calculated diagrams (see Fig.~\ref{LLoccupation}) and the
results of previous studies of similar
structures\cite{SchultzJCG98,SchultzPRB98} provides an important
input to the assignment of the measured absorption lines.


Sample B shows a single absorption line at low magnetic fields.
This line can be identified as a cyclotron resonance absorption
due to 2DEG confined in the QW, see Figs.~\ref{LLoccupation}b,
\ref{SPKT}b and \ref{FanChart}b.
The CR-like absorption is at low fields nearly linear with $B$ and
implies $m^{*}=0.022m_0$  for the carrier effective mass at the
Fermi level. Characteristically for narrow-gap materials with
strongly non-parabolic bands, the CR energy versus $B$ dependence
appears to be sub-linear at higher fields. Once the CR line
emerges above the reststrahlen band of HgTe/HgCdTe, it corresponds
to inter-LL transition $\delta$ (HH$_1$: $N=1
\rightarrow$ HH$_1$: $N=2$). Around $B\approx 7$~T, the $\gamma$
line (HH$_1$: $N=-1 \rightarrow$ HH$_1$: $N=0$) appears in the
spectrum and it gains in intensity at the expense of the $\delta$
line. At even higher fields, the $\gamma$ line weakens and finally
disappears from the spectrum at $B\approx17$~T. Assuming that
$B=17$~T corresponds to the Landau level filling factor $\nu=1$, we
find an excellent agreement upon the density of the 2DEG, which
has been derived from magneto-transport measurements on a parent
sample. Consistently, the $\alpha$ line (E$_1$: $N=0 \rightarrow$
HH$_1$: $N=1$) appears in the spectrum around $B\approx 6$~T, i.e.
for $\nu \leq 3$, and reaches its maximum intensity for $\nu=2$ at
$B\approx 12$~T. Its oscillator strength starts to gradually
decrease once filling factor decreases below $\nu = 1$.


The exact level of doping in sample A is unknown and therefore the
analysis of the magneto-absorption data of this sample is more
difficult. The appearance of the $\alpha$ transition is nevertheless
clear. This line is a common feature observed in both samples and
its energy position correlates well with the results of
calculations. In sample A, the $\alpha$ line remains visible in
the spectra up to magnetic fields of $B\approx 13$~T, which points
towards the residual $n$-type doping of this QW structure. This
doping level is however low, thus no CR-like transitions are seen.
The identification of the second line visible in the spectra of
sample A is a more subtle issue. Very likely, this line is
associated with the transition HH$_1$: $N=-2 \rightarrow$ HH$_1$:
$N=-1$, i.e. with the $\beta$ transition of
Fig.~\ref{LLoccupation}a. Nevertheless, the measured spectral
position of this line somewhat deviates from the theoretical
expectation and therefore we cannot exclude that it also involves
some over transitions which originate from LLs of the E$_1$ and/or
HH$_2$ subbands. In addition, the exact spectral shape of this
line is masked by the overlapping extra feature (appearing in both
sample at $\approx$50~meV) due to acceptor states in CdTe substrate. Notably,
in sample B, the $\beta$ transition is blocked at low magnetic
fields by $n$-type doping but surprisingly it does not appear in the
spectra at high magnetic fields $B>8$~T when $\nu<2$, i.e., when
the final-state LL associated with this transition is emptying. We
may speculate that the oscillator strength of the $\beta$ line is
reduced at high magnetic fields due to merging of the $N=-2$ LL
with the quasi-continuum of valence band LLs (see
Fig.~\ref{LLoccupation}).

\begin{figure}
\scalebox{0.68}{\includegraphics*{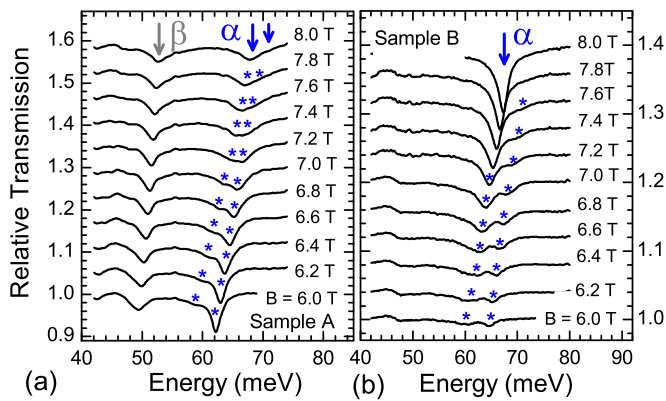}}
\caption{\label{Detail} Detail of magneto-transmission spectra for
samples A and B in parts (a) and (b), respectively, which show two
components of the $\alpha$ line at magnetic fields close to the
crossing field $B_c$.}
\end{figure}

Obviously, the above approach to interpreted the data does not
take into account all possible effects which may influence the
measured spectra. Among those are the effects of bulk inversion
symmetry and electron-electron interactions. Notably, these latter
effects may also largely influence the energies and degeneracy of
inter Landau transitions as we deal with strongly non-parabolic
systems for which Kohn's theorem does not hold.

In the following, our attention is focused on the $\alpha$ transition
which is an unambiguously defined spectral feature observed in
both investigated samples and involves the $N=0$ zero-mode LL in
the initial state. As can be seen in Fig.~\ref{Detail}, the $\alpha$ line 
has a character of a doublet visible in the spectra in a relatively narrow
range of magnetic fields, $B=6-8$~T. Exactly in this field
interval, at $B\approx7$~T, the crossing of the zero mode, $N=0$ and $N=-2$,
LLs is inferred from our calculations (see
Fig.~\ref{LLoccupation}). Instead, the observation of the doublet
structure indicates that those two LLs are never degenerate. Level
anti-crossing is one possible scenario to account for this
experimental fact. 

To model such an effect we have followed a
standard perturbation approach and phenomenologically
hybridized the initially calculated uncoupled $N=0$ and $N=-2$
LLs. The avoided crossing of the resulting coupled modes is shown
with dashed lines in Fig.~\ref{FanChart}. Assuming $\Delta=2$~meV
for the coupling parameter we roughly reproduce the observed
$\approx$4~meV~$(=2\Delta)$ energy separation between the doublet
components of the $\alpha$ line in the sample B. Doublet splitting seems 
to be slightly smaller in the sample A. We note, however, that the
appearance of the doublet structure in sample B is pretty abrupt
whereas our simple anti-crossing model would imply that this doublet
persists in somewhat wider range of magnetic fields. As discussed by
K\"{o}nig \emph{et al.},\cite{KonigScience07} the hybridization
of heavy-hole-like $\Gamma_8$ and $\Gamma_6$ states with opposite
spin direction and therefore coupling between the $N=0$ and $N=-2$
LLs may results from the breaking of bulk inversion symmetry,
initially neglected in our calculations. The theoretically
estimated amplitude of coupling parameter imposed by the
BIA terms reaches $\Delta\sim1.6$~meV in structures similar to
ours.\cite{KonigJPSJ08} 

Optionally, we speculate that the
observed splitting of the otherwise degenerate zero mode LLs at
the critical field $B_c$ might be also induced by
electron-electron interactions. We note that physics of our
zero-Landau level in the vicinity of $B_c$ might be similar to
that of a widely discussed case of the $n=0$ Landau level in
graphene. The degeneracy of this latter level is indeed believed
to be lifted by electron-electron interactions due to, for
example, spontaneous spin polarization.\cite{JungPRB09} Independently of its
origin, the persistent splitting between our zero mode LLs may
have a relevant consequences on the detailed nature of the
magnetic field driven insulator-metal-insulator transition. For
example, the metallic phase might be absent in this transition if
the system is kept as perfectly neutral. The accurate determination of the
BIA terms may also help to uncover more details of the QSHE, for
example, in reference to its quenching by the application of the
perpendicular magnetic field.

\section{Conclusions}

We have explored the fine structure of Landau levels in
[001]-oriented HgTe/HgCdTe quantum wells with an inverted band
structure using infrared magneto-spectroscopy. Our particular
attention has been focused on the low-lying conduction band and
top-most valence band Landau levels, i.e. on the zero-mode levels of
this Dirac-type system. We have found a spectroscopic signature of
avoided crossing (splitting at critical field $B_c$) of these levels.
This effect is likely due to the breaking of bulk inversion
symmetry, but perhaps also mediated by electron-electron
interactions. Our experimental findings may have a relevant
consequences on the subtleties of the physics of QSHE and magnetic-field-driven 
insulator-metal-insulator, the two characteristic and
intriguing phenomena observed in HgTe structures.

\begin{acknowledgments}
We acknowledge enlightening discussion with S. Wiedmann. This work
has been supported by EuroMagNET II under the EU contract
No.~228043, MSM0021620834, GACR No.~P204/10/1020, Barrande
No.~19535NF as well as DFG grants AS327/2 and MO771/12, the latter is part of a 
Japanese-German spintronics collaboration.
\end{acknowledgments}


\end{document}